\documentclass[review,12pt,3p]{elsarticle}

\usepackage{amssymb}
\usepackage{amsmath}

\usepackage{bm}
\usepackage{booktabs}
\usepackage{multirow}
\usepackage{makecell}
\usepackage{subfigure}
\usepackage{xcolor}
\usepackage{threeparttable}

\usepackage{hyperref}
\hypersetup{colorlinks=true,
            linkcolor=black,
            anchorcolor=black,
            citecolor=black,}
        
\usepackage[formats]{listings}
\usepackage{lstlinebgrd}
\lstdefineformat{Java}
{
  \{=\newline\string\newline\indent,%
  \}=\newline\noindent\string\newline,%
  ;=[\ ]\string\space,%
}
\usepackage{color}
\definecolor{dkgreen}{rgb}{0,0.6,0}
\definecolor{gray}{rgb}{0.5,0.5,0.5}
\definecolor{mauve}{rgb}{0.58,0,0.82}
\begin{document}

\begin{frontmatter}



\title{Revisiting Evolutionary Program Repair via Code Language Model}


\author[buaa]{Yunan Wang}
\ead{21373161@buaa.edu.cn}

\author[buaa]{Tingyu Guo}
\ead{tingyuguo@buaa.edu.cn}

\author[buaa]{Zilong Huang}
\ead{21371394@buaa.edu.cn}

\author[buaa]{Yuan Yuan\corref{corresponding}}
\ead{yuan21@buaa.edu.cn}
\cortext[corresponding]{Corresponding author.}
\affiliation[buaa]{organization={School of Computer Science and Engineering, Beihang University},
            city={Beijing},
            country={China}}

\begin{abstract}
Software defects are an inherent part of software development and maintenance. To address these defects, Automated Program Repair (APR) has been developed to fix bugs automatically. With the advent of Large Language Models, Code Language Models (CLMs) trained on code corpora excels in code generation, making them suitable for APR applications. Despite this progress, a significant limitation remains: many bugs necessitate multi-point edits for repair, yet current CLM-based APRs are restricted to single-point bug fixes, which severely narrows the scope of repairable bugs. Moreover, these tools typically only consider the direct context of the buggy line when building prompts for the CLM, leading to suboptimal repair outcomes due to the limited information provided. This paper introduces a novel approach, ARJA-CLM, which integrates the multiobjective evolutionary algorithm with CLM to fix multilocation bugs in Java projects. We also propose a context-aware prompt construction stratege, which enriches the prompt with additional information about accessible fields and methods for the CLM generating candidate statements. Our experiments on the Defects4J and APR-2024 competition benchmark demonstrate that ARJA-CLM surpasses many state-of-the-art repair systems, and performs well on multi-point bugs. The results also reveal that CLMs effectively utilize the provided field and method information within context-aware prompts to produce candidate statements. 
\end{abstract}




\end{frontmatter}



\section{Introduction}
Software defects are inevitable during the development and maintenance processes, which has caused huge financial losses. To repair these defects, developers need to spend about 50\% of the overall development time \cite{b23}. Automated Program Repair (APR) is promising to reduce the manpower of repairing defects in software.

Automated program repair \cite{b23,b34} aims to find defects in programs and try to give semantically correct fixes so that programs can run correctly. Heuristic search is widely used in this field. Heuristic-based APR first identifies multiple Likely Buggy Statements (LBSs) with fault localization \cite{b46,b47,b48,b49,b50,b51}. For each LBS, it prepares a set of candidate statements for modification. Finally it employs heuristic algorithms \cite{b16} to search for a series of modifications that would allow the revised source code to pass the corresponding test cases. Although such repair tools have proven effective for fixing multilocation defects, their success heavily relies on the quality of candidate statements. Consequently, much research is dedicated to developing methods for generating high-quality candidate statements. Among them, ARJA-e (Automated Repair of Java Programs based on Genetic Programming) \cite{b1,b29,b54}, a efficient heuristic-based repair approach, leverages the statement redundancy assumption \cite{b33} and the repair templates \cite{b37} to extract candidate statements from other parts of the program and generate statements based on artificially defined templates respectively. However, these methods are too templated and programmed, which cannot make good use of the comprehensive semantic information of the LBS context, resulting in a low probability of the search space containing the patches. 

\raggedbottom
With the rapid development of Code Language Models (CLMs), they have shown good performance in code generation and code completion tasks, and have been preliminarily applied in the field of APR \cite{b2,b30,b31}. Although the current CLM-based repair tool has a good repair effect, it can only repair single-point defects, which greatly limits the range of repairable defects. In addition, these tools only leverage the original LBS and their direct context to construct the prompts for CLM, which restricts the CLM from obtaining defect-related context information and generating statements with callable methods and fields within the class where the LBS is located.

\vspace{0.1cm}
\noindent \textbf{Our Work}. To solve the above problems, we propose ARJA-CLM. Different from previous approaches using language models \cite{li2022dear} or tree distance algorithms \cite{saha2019harnessing} to evaluate the similarity of LBSs for locating multi-point defects, we revisit the multiobjective evolutionary algorithm to search for multilocation defects, and combine it with the advances of CLMs. Consequently, on the one hand,  the introduction of the multiobjective evolutionary algorithm expands the range of repairable defects of the CLM-based repair tool; on the other hand, ARJA-CLM can leverage the high-quality candidates generated by CLMs to efficiently search for patches. Additionally, we propose a new method of constructing context-aware prompt for each LBS. The key insight is that besides the direct context of the LBS, ideally, the code model needs to be aware of the entire defective project and its callable components to generate the correct fill-in statements. However, due to the limited context length of language models, it is nearly impossible to input the entire project into the model. Therefore, to help CLMs perceive the relevant contextual information and make full use of the model’s context window, we innovatively add the callable field and method information of the LBS into the prompt. While ARJA-CLM is generalizable to diverse CLMs, we employ the latest released CodeLlama \cite{b11} and InCoder \cite{b12} to generate candidate statements. Specifically, we first use fault localization to locate the LBSs. Then we extract the context as well as the callable fields and methods of the LBS position to construct the prompts and input them into the model with a series of candidate statements returned. Finally, we use the multiobjective evolutionary algorithm to search these candidate statements for plausible patches. In addition, we try to combine the candidate statements generated by ARJA-e with those output by CLM to explore the possibility of improving the repair effect by expanding the search space. We also study the repair performance of ARJA-CLM using different sizes and types of models to investigate the relationship between model performance and repair effect.

Following previous work, we conduct experiments on 224 real-world defects in the Defects4J benchmark. Experimental results show that ARJA-CLM fixes 64\% more defects compared to ARJA-e and 43\% more defects compared to the latest learning-based multi-point defect repair tool DEAR \cite{li2022dear}. Additionally, it fixed 22 defects that other tools could not, demonstrating the great potential of combining cutting-edge CLMs with traditional repair methods. Furthermore, comparative experiments show that the context-aware prompts successfully improved repair effectiveness of ARJA-CLM by 21\%. 

To further verify the generalizability of ARJA-CLM, we also conduct experiments on the newly released APR-2024 competition dataset \cite{shariffdeen2024program} to mitigate the impact of dataset leakage. The results show that compared to ARJA-e, ARJA-CLM generated 14 (280\%) more plausible patches, surpassing all participating repair tools, demonstrating ARJA-CLM’s superiority in error algorithm solution repair.

To sum up, this paper makes the following contributions:
\begin{itemize}
\item \textbf{New Dimension}. We bridge the gap between CLM-based APR and heuristics-based APR, which employs CLMs to obtain a higher-quality search space and use the heuristic algorithm to search for patches. Different from current CLM-based APR focusing on fixing single-point defects, our approach enables CLMs to repair multi-point defects, which is closer to real-world situations. Notably, we demonstrate the potential of using CLMs to address the candidate statement preparation problem for the heuristic-based APR.
\item \textbf{Technique}. We propose ARJA-CLM, which applies the recently released pre-trained model CodeLlama \cite{b11} and InCoder \cite{b12} to APR by predicting the correct statement in masked buggy position with their powerful code-filling capabilities without any further training or fine-tuning. To improve the search efficiency, we revisit the multiobjective evolutionary algorithm proposed by ARJA-e \cite{b1} to search the candidate statements for patches. More importantly, since ARJA-CLM’s candidate generation and search phases are separate, it can easily generalize to various CLMs for code filling in practice.
\item \textbf{Context-aware Prompt Construction}. We propose a new method of constructing prompts for LBSs to help CLMs aware of the whole defective project, which adds the callable fields and methods of the defective location to the prompts, thus enabling the CLM to leverage them for generating code sequences. Experimental results show that context-aware prompts improve ARJA-CLM’s repair efficiency by 21\%.
\item \textbf{Extensive Study}. We conduct large-scale experiments on ARJA-CLM. The experimental results on the widely-adopted Defects4J benchmark show that ARJA-CLM demonstrates superior defect-fixing capabilities by fixing 64\% more defects than ARJA-e and 43\% more than DEAR, a leading learning-based tool. What's more, ARJA-CLM surprisingly fixs unique 22 defects that other tools couldn't. Its repair capability is further confirmed on the newly released APR-2024 competition benchmark, where it generated 14 (280\%) more plausible patches than ARJA-e, outperforming all other participating repair tools. Meanwhile, additional study are conducted to explore the possibility of improving the repair effect by expanding the search space through combining candidate statements from multiple sources and to explore the impact of model types and sizes on the repair performance of ARJA-CLM.
\end{itemize}

\section{Background}
\subsection{Automated Program Repair}
Automatic Program Repair (APR) is a technique that uses program analysis techniques to automatically detect and fix bugs or vulnerabilities in software. APR can improve the quality and security of software, reduce the debugging burden of developers, and save the cost and time of software maintenance. 

The current defect repair methods are mainly divided into traditional repair approaches and learning-based repair approaches. The former can be subdivided into heuristic-based approaches \cite{b52,b44,b22}, template-based approaches \cite{b37,b69}, constraint-based approaches \cite{b70,b71}, etc. Among them, the heuristic-based APR uses heuristic algorithms to search and generate patches. Heuristic algorithms, such as genetic algorithms and simulated annealing, are a class of algorithms that can find approximate optimal solutions in complex search spaces. heuristic-based APR has strong generality, especially in the repair of multilocation defects. The search space obtained by the current heuristic-based APR has a low probability of containing patches. To solve this problem, ARJA-CLM uses the CLM to better understand the semantic information of the defect location context and generate higher-quality candidate statements to modify the LBS.

Learning-based APR uses deep neural networks to repair software defects. This technique uses a large amount of open-source code and defect data to learn the syntax and semantic features of code, thus improving the accuracy and efficiency of repair. Learning-based APRs are mainly divided into two categories: based on sequence-to-sequence generation and based on the pre-trained code language model. To distinguish them, in the following content, learning-based APR specifically refers to those based on sequence-to-sequence generation. In this paper, we focus on CLM-based APR, which leverages prior knowledge learned from large-scale code corpus to predict the missing or buggy parts in the defective location. However, the current CLM-based \cite{b56,b2} APR can only repair single-point defects and focuses the repair effect under perfect fault localization setting, which deviates from the real-world repair situation. To solve these problems, ARJA-CLM introduces the multiobjective evolutionary algorithm and gives the experimental results under not perfect fault localization setting.
\subsection{Evolution-based APR}\label{evolution}
Genetic programming is widely used in software engineering \cite{b18}. Since the heuristics-based APR GenProg \cite{b14,b22,b44} which uses genetic programming was proposed, many researchers have proposed various APR based on evolutionary algorithms. This paper uses the multiobjective evolutionary algorithm by ARJA-e \cite{b1}, which is based on NSGA-II \cite{b17}. We take this algorithm as an example to introduce the evolutionary algorithm in APR. The following are divided into three parts: patch representation, fitness function, and gene operation.
\subsubsection{Patch Representation}
To encode patches as genomes, ARJA-e uses $\bm{x=(b,u,p,q)}$ to represent a patch, where $\bm{b}$, $\bm{u}$, $\bm{p}$, $\bm{q}$ are both $n$-dimensional vectors and $n$ represents the max number of LBSs of the patch. For each LBS, the patch may delete the LBS, insert a candidate statement before the LBS, replace the LBS with a candidate statement, or do nothing. Let $x_j(1\leq j \leq n)$ represent the operation on the $j^{th}$ LBS. $b_j$ takes values of 1/0, respectively indicating whether to modify/not to modify the $j^{th}$ LBS. $u_j$ takes values of 1/2/3, respectively indicating whether to delete/replace/insert before the LBS. $p_j$ takes a specific value to indicate selecting the $p_j^{th}$ statement in candidate replacement set to replace the LBS. $q_j$ takes a specific value to indicate selecting the $q_j^{th}$ statement in candidate insertion set to insert before the LBS.
\subsubsection{Fitness Function}
To evaluate the fitness of each patch $\bm{x}$ and select the best individuals in the population, ARJA-e defines two fitness functions. One of them is $f_1(\bm{x})$ as follows, which aims at minimizing the number of effective modifications.
$$f_1(\bm{x})=\sum_{j=1}^n b_j$$

The second fitness function aims to minimize the failure rate. Since Java development relies on the JUnit framework as the main test suite, ARJA-e considers the JUnit assertions (denoted by $e$) in each test point, such as \textbf{assertEquals(double expected,double actual,double delta)} indicating that if the program runs correctly, the actual running result (denoted by $x$) and the expected running result (denoted by y) should be equal within the error of delta (denoted by $\delta$). Therefore, ARJA-e defines the gap between the actual and the expected of each assertion as follows, where $v(x)$ is a normalizing function in [0, 1].
$$
d(e) = \begin{cases}
\begin{array}{ll}
v(|x-y|-\delta), & |x-y| \geq \delta \\
0, & |x-y| < \delta
\end{array}
\end{cases}
$$

Since a JUnit test point (denoted by $t$) consists of many assertions, and there are loops and branches in the test point, which affect the actual number of assertions that run during the test process, ARJA-e defines $E(\bm{x},t)$ as the set of all assertions that actually run on test point $t$ after the program is modified by patch $\bm{x}$ and defines $h(\bm{x},t)$ as the failure rate on $t$ as follows. 
$$h(\bm{x},t)=\frac{\sum\nolimits_{e\in E(\bm{x},t)}d(e)}{|E(\bm{x},t)|}$$

Let $T_{pos}$ denotes the set of test points that patch $\bm{x}$ passes, and $T_{neg}$ denotes the set of test points that $\bm{x}$ fails, then the function $f_2(\bm{x})$ representing the failure rate of $\bm{x}$ is as follows, where $w$ is a predefined parameter.
$$f_2(\bm{x})=\frac{\sum\nolimits_{t\in T_{pos}}h(\bm{x},t)}{|T_{pos}|}+w\times \left(\frac{\sum\nolimits_{t\in T_{neg}}h(\bm{x},t)}{|T_{neg}|}\right)$$
\subsubsection{Gene Operation}
Gene operations include crossover and mutation, which are used to obtain offspring populations from parent populations. For crossover operation, ARJA-e uses the general HUX (half uniform crossover) crossover operator for the four parts of the parent patch $\bm{x=(b,u,p,q)}$. For mutation operation, ARJA-e first selects an LBS modification $(b_j,u_j,p_j,q_j)$ of the initial offspring patch $\bm{x}$, then respectively performs flip mutation on $b_j$, and performs uniform mutation which randomly selecting a value from the possible values on $u_j$,$p_j$,$q_j$.
\subsection{Large Pre-trained Language Model}
In recent years, various Large Language Models (LLMs) \cite{b32,b39,b40,b41} have been widely used in many fields. LLMs accept prompts as input and give corresponding answers based on the context information. A LLM is composed of a series of Transformer \cite{b25} modules (including encoder and decoder modules) stacked together. The encoder is responsible for encoding the input sequence into a hidden vector, and the decoder is responsible for predicting the next word for the target sequence based on the hidden vector and the generated content. Transformer modules are composed of attention layers, which enable the model to understand the connections between the words that make up the sentence and focus on the relevant parts when processing the input data. Large models use a huge amount of corpus data from the Internet web pages for training. For a given prompt sentence, LLM first divides and encodes it into a word vector sequence, and uses position embedding (e.g., RoPE \cite{b26}) to add the position information of each word to the word vector. After a series of Transformer module calculations, the probability of each word in the vocabulary appearing in the next or masked part of the sentence is obtained, and the word is output by sampling.

Code language models \cite{b11,b12,b13,b42} are LLMs trained on vast code corpora. Given a piece of code with a blank, CLMs can predict the blank based on the context information. Current CLMs perform remarkably well in the code completion task, which makes them very suitable for repairing defects by predicting the correct statement based on the context information of the buggy position. In this paper, we use the recently released model InCoder and CodeLlama to generate candidate statements.

\section{Approach}
\begin{figure}[htbp]
\centerline{\includegraphics[width=\linewidth]{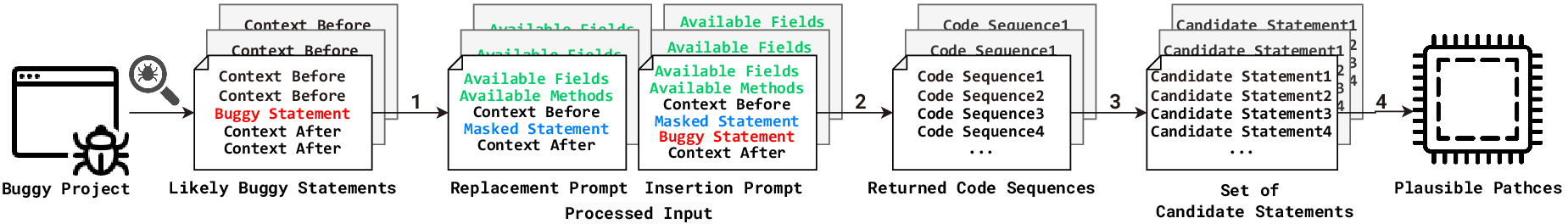}}
\caption{ARJA-CLM overview}
\label{overview}
\end{figure}

In this section, we propose a new APR that combines CLM and multiobjective evolution to generate patches. It uses CLM to generate candidate statements through the cloze-style method \cite{b2} which masks the buggy line and queries CLM to fill the blank given its surrounding context. These candidate statements will be selected for replacing or inserting in front of LBS.  Then it uses the multiobjective evolutionary algorithm in section \ref{evolution} to search for plausible patches from the candidate statements.

We also design a context-aware prompt for each LBS. Except for contextual information, we add to the prompt the callable fields and methods of the class where the LBS is, so that the CLM can leverage them to generate candidate statements. Figure \ref{overview} provides an overview of our approach:
\begin{itemize}
\item \textbf{Step 1 (Section \ref{prompt build})}: For each defective project, we get several LBSs by fault localization. For each LBS, we extract the context before and after the LBS, callable fields and methods of the class where the LBS is, and the original LBS. Then we leverage them to build context-aware prompts as CLM input. 
\item \textbf{Step 2 (Section \ref{sequence generate})}: For each prompt, we encode it with a tokenizer and input the token representation to CLM with the return of several code sequences. It should be mentioned that to expand the search space, we configure the sampling settings to maximize the randomness of the sentence outputs. Moreover, we exploit the parallelism of GPU to generate several code sequences at once to improve output efficiency. 
\item \textbf{Step 3 (Section \ref{set built})}: For each LBS, we convert the code sequences into Java statement components, and add them to the set of candidate replacement and insertion statements. 
\item \textbf{Step 4 (Section \ref{patches search})}: Now that for each LBS, we have prepared several candidate statements for replacing or inserting in front of it. We find plausible patches by multiobjective evolution.
\end{itemize}

\subsection{Build Context-Aware Prompt for CLM Input}\label{prompt build}
\begin{figure}[h]
\centerline{\includegraphics[width=0.8\linewidth]{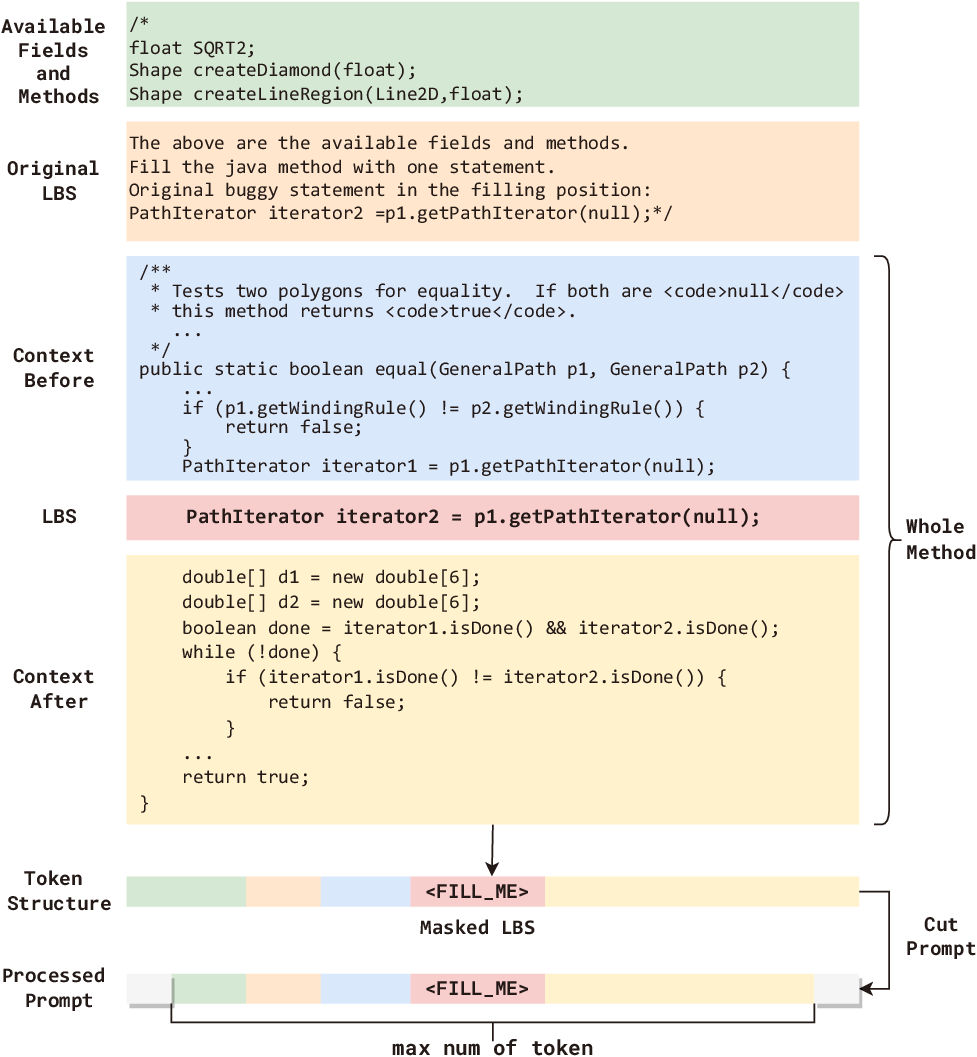}}
\caption{Example of processed context-aware replacement prompt}
\label{prompt}
\end{figure}

For each buggy project, we use a fault localization technique called Ochiai \cite{b3,b4} to locate a list of LBSs. We consider at most $n_{max}$ LBSs and ignore these LBSs with $susp$ (indicating the faulty probability of the LBS) lower than a threshold $\gamma_{min}$. We build two prompts for each LBS, corresponding to update of replacement and insertion respectively. Figure \ref{prompt} shows the structure of the processed context-aware replacement prompt:
\begin{itemize}
\item \textbf{Step 1}: We extract the whole buggy method where the LBS is located. Since we have modification methods of replacing or inserting before LBS, we replace the LBS with a special token ( e.g., $<$FILL\_ME$>$ ) or insert the token before the LBS respectively to mark the blank for CLM. In addition, for the modification method of replacing the LBS, we place the original LBS as a comment before the buggy method.  
\item \textbf{Step 2}: We extract the callable fields and methods of the class where the LBS is located and put them at the start of the prompt in the form of comments. For each callable field, we provide the type and name. For each callable method, we provide its return type, name, and parameter types.
\item \textbf{Step 3}: Due to the maximum context length limit and computing resource limit, the excessive length of the buggy method and extra callable fields and methods will make the prompt too long, thus preventing the large model from generating code sequences. Therefore, we trim the prompt. We start from the position of the LBS and expand the context until the number of tokens reaches a threshold $max_{tokens}$.
\end{itemize}
\subsection{Generate Code Sequences}\label{sequence generate}
Now that we have prepared prompts for each LBS,  we use the tokenizer of CodeLlama and InCoder to encode it respectively. The tokenizers of both models use the BPE word segmentation algorithm \cite{b27} involved in SentencePiece \cite{b28}, which encodes rare and unknown words into subword sequences and reduces the size and computation of the vocabulary. We input the token representation to CLM to generate multiple code sequences in parallel. Moreover, we enable model sampling to obtain diversified code sequences as randomly as possible. Table \ref{sampling setting} shows the sampling parameter setting.

\begin{table}[htbp]
\caption{Sampling setting}
\resizebox{\textwidth}{!}{
\begin{tabular}{clc} 
\toprule 
Parameter & Description & Value \\ 
\midrule 
top\_p & Model selects from tokens whose cumulative probability exceeds $p$ from the output distribution. & 0.9 \\
\midrule 
top\_k & The model only samples from the top $k$ tokens with the highest probability. & 50 \\
\midrule 
temperature & Indicates the randomness of sampling with the value range of [0,1]. & 1.0 \\
\midrule 
num\_return\_sequences & The number of returned sequences at once. & 10\\
\midrule 
max\_new\_tokens & The max length of the token representation of returned sequences. & 100 \\
\bottomrule 
\end{tabular}}
\label{sampling setting}
\end{table}
\subsection{Build the Set of Candidate Statements}\label{set built}

\begin{figure}[h]
\centerline{\includegraphics[width=0.9\linewidth]{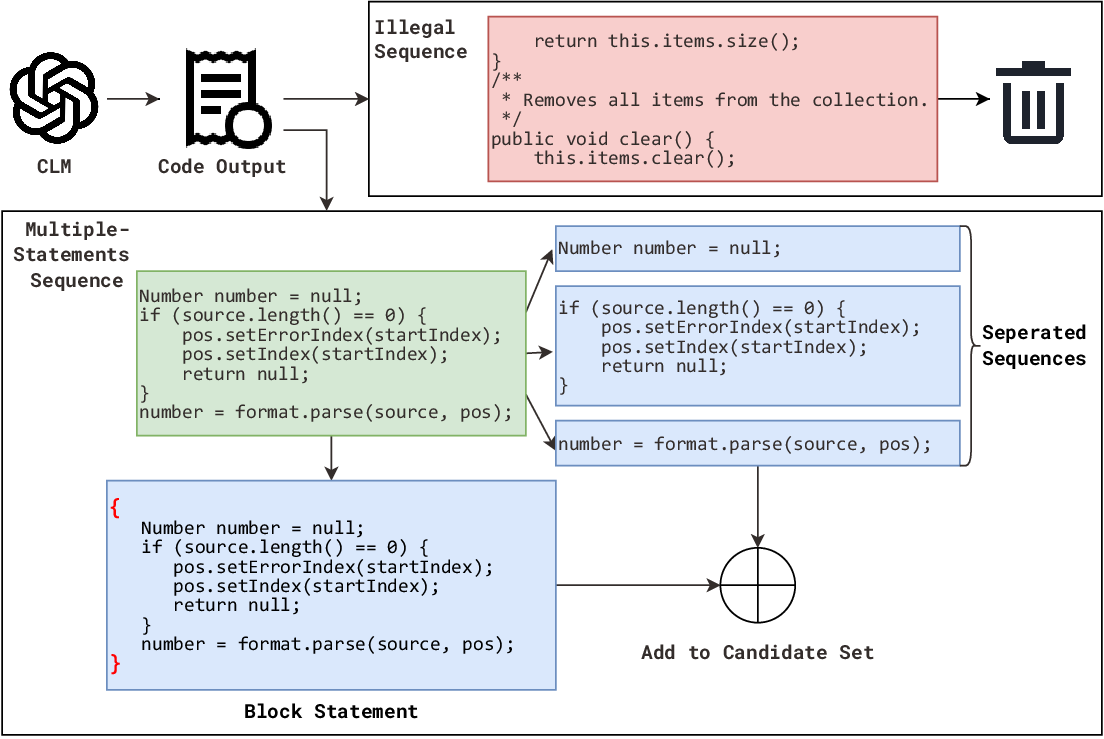}}
\caption{Example of code sequence transformation}
\label{transformation}
\end{figure}

We convert each code sequence from CLM to a Java statement element and add it to the candidate set. Since the code sequence output by the model may not be a valid statement element, we filter out these erroneous code sequences. For the case where the code sequence output by the model contains multiple statement elements, we treat the sequence as a block statement that consists of several statements and add it to the candidate set; in addition, to expand the search space, we split it into several independent statement syntax elements and add all of them to the candidate statement set. The specific transformation example is shown in Figure \ref{transformation}.

To simplify and optimize the search space, we remove duplicate statements and statements identical to the original LBS from the set of candidate statements.

\subsection{Search for Patches by Multiobjective Evolution}\label{patches search}
Now for the buggy project, we have LBSs that may contain faults. For each LBS, we have two sets of candidate statements which will be selected respectively to replace or insert before the LBS to fix it. Then we perform multiobjective evolution in Section \ref{evolution} to search for plausible patches.  To compare repairing performance with the original ARJA-e, the algorithm parameters remain the same as in ARJA-e, as shown in Table \ref{evolotion setting}.

\begin{table}[!htbp]
\centering 
\caption{Multiobjective evolution setting \cite{b1}} 
\renewcommand{\arraystretch}{0.7}
\begin{tabular}{cp{8cm}c} 
\toprule 
Parameter & Description & Value \\ 
\midrule 
$N$ & Population size & 40 \\
\midrule 
$G$ & Maximum number of generations & 50 \\
\midrule 
$\gamma_{min}$ & Threshold for the suspiciousness & 0.1 \\
\midrule 
$n_{max}$ & Maximum number of LBSs considered & 60\\
\midrule 
$w$ & Weights of fitness function $f_2$ in Section \ref{evolution} & 0.5 \\
\midrule 
$\mu$ & Weight to initialize the initial population & 0.06 \\
\bottomrule 
\end{tabular}
\label{evolotion setting}
\end{table}
\section{Experimental design}
\subsection{Research Questions}
In this paper, we study the following research questions:
\begin{itemize}
\item \textbf{RQ1}: How does the type and size of the model affect the quality of the candidate statements from CLM?
\item \textbf{RQ2}: Which source of search space has a better repair effect?
\item \textbf{RQ3}: Does the context-aware prompt contribute to better repair performance?
\item \textbf{RQ4}: Compared to ARJA-e, how does ARJA-CLM perform on multilocation bugs?
\item \textbf{RQ5}: How does ARJA-CLM compare against state-of-the-art APR approaches?
\end{itemize}
To answer RQ1, we use the latest open-source CLMs CodeLlama-7B, InCoder-1.3B, and InCoder-6.7B to generate candidate statements. We compare CodeLlama-7B and InCoder-6.7B, two models of similar size but different types, and InCoder-1.3B and InCoder-6.7B, two models of the same type but different sizes. To answer RQ2, we conduct experiments respectively on the candidate statements from CodeLlama, the original ARJA-e, and their combination. To answer RQ3, we provide the CLM with prompts with and without extra field and method information and obtain two sets of candidate statements respectively before the search. To answer RQ4, we count the number of multilocation bugs plausibly and correctly fixed by ARJA-CLM and ARJA-e. To answer RQ5, we compare the repair effect of ARJA-CLM with DEAR \cite{li2022dear}, HERCULES \cite{saha2019harnessing} which also aims at repairing multilocation bugs; we also choose state-of-the-art repairing tools such as  Recoder \cite{b5}, DLFix \cite{b6} based on machine learning, AlphaRepair \cite{b2}, GAMMA \cite{zhang2023gamma} based on CLMs and TBar \cite{b7}, SimFix \cite{b8} based on traditional methods for comparison.
\subsection{Implementation}
Except for RQ3, the model prompts in the rest of the experiments all have extra field and method information. To cut the inference cost, all models perform inference in half-precision (i.e., representing model weights with 16-bit floating point numbers). Following ARJA-e, ARJA-CLM  uses not perfect fault localization, which depends on the suspicious locations detected by Ochiai \cite{b3,b4}. All experiments are conducted on the DELL C4140 cluster running Ubuntu 22.04.3 LTS with CPU of Intel Xeon Gold 6148 CPU @ 2.40GHz, GPU of NVIDIA Tesla V100 32GB, and Java version of 1.7.0\_80. All experiments only perform one round of multiobjective evolution on the candidate statements, and the time of one round of search is within 1 hour.
\subsection{Dataset of Bugs}
\textbf{Defects4J}. For evaluation, we use a benchmark called Defects4J \cite{b21} that has been wildly used for evaluating Java repair tools. Following ARJA-e, we conduct experiments on Defects4J version v1.0.1 with 224 real-world bugs in four projects: Chart, Lang, Math, Time. We ignore Mockito because its compilation requires a complex compilation process. We ignore Closure because it uses the customized testing format instead of the standard JUnit tests. For each buggy project, Defects4J provides the source code that contains a bug, and an associated  JUnit test suite with at least one failed test case. Following previous work, we manually verify the correctness of the plausible patches by our repair approach. For each bug, as the evolutionary algorithm of ARJA-CLM has the objective of minimizing the number of modifications, we only select the smallest 10 patches for inspection and skip the overfitting patches (e.g., those deleting necessary statements).

\textbf{APRCOMP-2024 AI Generated Code Track (Java)} \cite{shariffdeen2024program}. We notice the possibility of source code from Defect4J leaking into CodeLlama which we uses. Therefore, to demonstrate that ARJA-CLM is less affected by this problem and verify the generalization of ARJA-CLM, we conduct experiments on the newly released ARPCOMP dataset. This dataset contains 100 buggy codes, each generated by OpenAI’s GPT-3.5 and GPT-4 models, aimed at solving general algorithm problems on Leetcode. Each code has its corresponding public test suite and private test suite, where the public test suite is visible to the repair tools and the private test suite is for verifying the correctness of the plausible patches generated by the repair tool. 

\section{Result analysis}
\subsection{Overview}
For each defect from Defects4J, we use CodeLlama-7B, InCoder-1.3B, InCoder-6.7B, and ARJA-e to generate candidate statements respectively and combine the statements from CodeLlama-7B and ARJA-e. Then we perform multiobjective evolution on the five sets of statements above. Note that the model prompts here all have extra method and field information. We get plausible patches, and the results are shown in Table \ref{result overview}.

\begin{table}[h]
\centering 
\caption{Result overview of plausible patches} 
\renewcommand{\arraystretch}{0.7}
\resizebox{\textwidth}{!}{\begin{tabular}{c|ccccc} 
\toprule 
Project & CodeLlama-7B & InCoder-6.7B & InCoder-1.3B & ARJA-e & CodeLlama+ARJA-e\\ 
\midrule 
Chart & 14 & 11 & 9 & 14 & 16 \\
Lang & 28 & 27 & 20 & 23 & 30 \\
Math & 46 & 31 & 30 & 36 & 42 \\
Time & 6 & 2 & 3 & 3 & 6\\
\midrule 
Total Plausible & 94 & 71 & 62 & 76 & 94 \\
\midrule
Plausible Percentage & 41.96\% & 31.70\% & 27.68\% & 33.93\% & 41.96\% \\
\bottomrule 
\end{tabular}}
\label{result overview}
\end{table}

Among the three models, we notice that ARJA-CLM using CodeLlama-7B can generate plausible patches for most defects. So we further examine the correct patches and show the plausible and correct patches generated by ARJA-CLM using CodeLlama-7B in Table \ref{CodeLlama}.

\begin{table*}[h]
\centering 
\caption{ List of the bugs fixed and correctly fixed by ARJA-CLM using CodeLlama-7B} 
\resizebox{\textwidth}{!}{\begin{tabular}{ccc}
\toprule 
Project & Plausible & Correct\\ 
\midrule 
\multirow{2}*{Chart} 
&\makecell[l]{
C1,C3,C4,C5,C7,C9,C10,C11,C13,C14,C18,C19,C24,C25}
&\makecell[l]{
C1,C3,C9,C10,C11,C14,C19,C24
}\\
                \cmidrule{2-3}
~&\multicolumn{1}{l}{$\sum=14$} & \multicolumn{1}{l}{$\sum=8$}   \\
\midrule 
\multirow{2}*{Lang} 
&\makecell[l]{
L6,L7,L8,L10,L13,L16,L20,L22,L24,L27,L28,L33,L35,L37,\\
L39,L40,L44,L45,L50,L51,L54,L55,L57,L58,L59,L61,L63,L64}
&\makecell[l]{
L7,L8,L10,L13,L20,L24,L28,L33,L35,L37,L39,L40,L45,L51,L54,\\
L57,L59,L64}\\
                \cmidrule{2-3}
~&\multicolumn{1}{l}{$\sum=28$} & \multicolumn{1}{l}{$\sum=18$}   \\
\midrule 
\multirow{2}*{Math} 
&\makecell[l]{
M2,M3,M4,M5,M6,M7,M8,M22,M24,M27,M28,M30,M32,M34,\\
M41,M42,M44,M46,M49,M50,M53,M57,M58,M59,M60,M63,M70,\\
M71,M73,M75,M77,M78,M79,M80,M81,M82,M84,M88,M89,M93,\\
M95,M98,M99,M101,M103,M105}
&\makecell[l]{
M2,M3,M4,M5,M22,M24,M27,M30,M34,M41,M50,M53,M57,M58,\\
M60,M63,M70,M75,M77,M80,M89,M95,M98,M99,M101,M103,M105}\\
                \cmidrule{2-3}
~&\multicolumn{1}{l}{$\sum=46$} & \multicolumn{1}{l}{$\sum=27$}   \\
\midrule 
\multirow{2}*{Time} 
&\makecell[l]{T4,T9,T11,T15,T20,T24}
&\makecell[l]{T9,T11,T15}\\
                \cmidrule{2-3}
~&\multicolumn{1}{l}{$\sum=6$} & \multicolumn{1}{l}{$\sum=3$}   \\
\midrule
Total & \makecell[l]{94 (41.96\%)} & \makecell[l]{56 (25\%) } \\
\bottomrule 
\end{tabular}}
\label{CodeLlama}
\end{table*}

\subsection{RQ1.Impact of the type and size of CLM}
To explore the impact of model size and type on repair effect, we use InCoder-1.3B, InCoder-6.7B, and CodeLlama-7B to generate candidate statements under the same conditions and search for plausible patches in these three candidate statement sets.

As shown in Table \ref{result overview}, we find that comparing InCoder-1.3B and InCoder-6.7B, the repair effect improves as the model parameter size increases. This is within expectations, as the larger the model parameter size, the more complex functions the model can represent, thus improving the generalization ability and accuracy of models in in-fill generations.

Comparing InCoder-6.7B and CodeLlama-7B, under the approximate parameter size, ARJA-CLM using CodeLlama-7B has a significantly better repair effect. We speculate that this may be due to the differences in overall model performance, which make CodeLlama superior to InCoder in generating candidate statements.

To prove our assumption, we give the scores of the two models in the HumanEval pass@1 test. CodeLlama-7B scores 33.5\% \cite{b11} while InCoder-6.7B scores 15.2\% \cite{b12}. HumanEval pass@1 \cite{b43} is a benchmark to evaluate code generation models. It only allows the model to generate code once for each problem, then uses unit tests to verify the correctness of the code, and finally calculates the percentage of problems that pass the test. It can be inferred from the score that CodeLlama-7B has a higher accuracy than InCoder-6.7B in generating candidate statements, thus making ARJA-CLM using CodeLlama perform better. 

\begin{figure}[htbp]
\centerline{\includegraphics[width=0.8\linewidth]{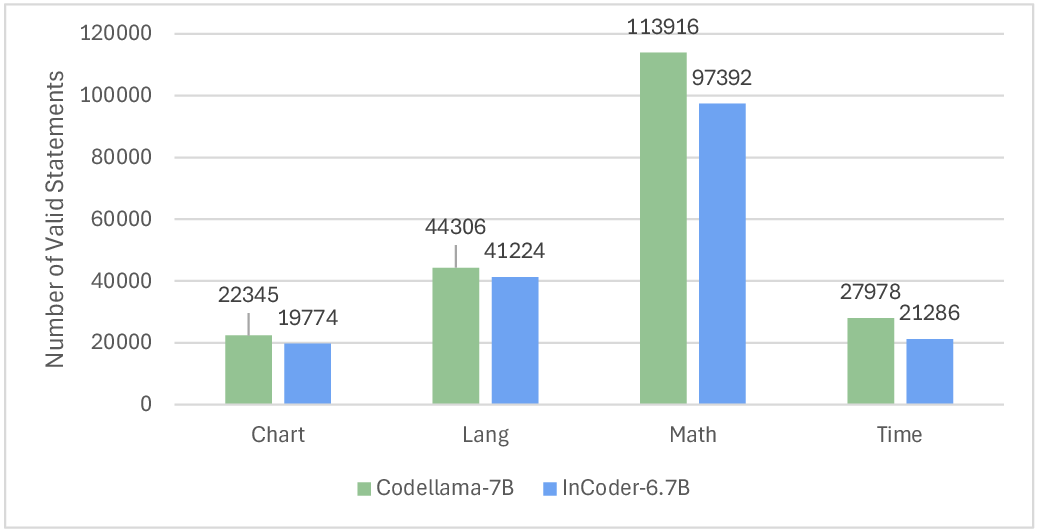}}
\caption{Number of valid statements from CodeLlama and InCoder}
\label{statementcnt}
\end{figure}

Here we further reflect the relationship between overall model performance and the ability to generate candidate statements in APR tasks. In the code completion task, if the model output does not end with \textless\textbar endofmask\textbar\textgreater\ or other end tokens, it means that the output is likely to be wrong or invalid, that is, the model cannot correctly complete the filling task for a certain input; in addition, the code sequence end with \textless\textbar endofmask\textbar\textgreater\  may also not be a valid and complete statement as shown in Figure \ref{transformation}, which indicates that the model is not suitable for the in-filling task on the statement level. Since the input prompts and output settings of models are consistent, after filtering out the above two types of invalid outputs, the total number of candidate statements reflects the ability of the model to generate candidate statements to some extent. So we count the total number of candidate statements generated by CodeLlama-7B and InCoder-6.7B on four projects. As shown in Figure \ref{statementcnt}, CodeLlama-7B can generate more valid statements than InCoder-6.7B on four projects, thus showing the superiority of CodeLlama-7B in the repair task with statement granularity. It demonstrates that better model performance leads to better candidate statement generation for APR tasks.

\subsection{RQ2.Impact of different sources of search space}
To investigate the impact of different sources of search space on the repair effect, we perform multiobjective evolution on three search spaces (i.e., set of candidate statements): generated by CodeLlama, generated by ARJA-e based on the statement-level redundancy assumption and repair templates, and the combination of both.

As shown in Table \ref{result overview}, search space from CodeLlama and CodeLlama+ARJA-e can both fix 94 bugs, significantly higher than those from ARJA-e alone. It proves the superiority of the CLM over the traditional method based on the statement-level redundancy assumption and repair templates in terms of generating candidate statements. 

\textbf{The impact on the final repair effect}. To further distinguish the repair effect, we count the number of correct patches for these three search spaces. As shown in Table \ref{differentsource},  space from CodeLlama can output more correct patches than CodeLlama+ARJA-e (i.e., combination of search space from both CodeLlama and ARJA-e), thus indicating that combining the search spaces before searching does not improve the repair effect. We infer that although combining search space from different sources can expand the search space and increase the diversity of candidate statements, considering that the number of candidate sentences for each LBS is relatively large, combining space also exponentially increases the difficulty for the evolutionary algorithm to find the correct patch. We also notice that the correct repair rate of the space from CodeLlama+ARJA-e is very close to the average of the repair rates of CodeLlama and ARJA-e. We speculate that this is because the two search spaces have different probabilities of containing patches, and mixing the two search spaces when they are fairly complex leads to the averaging of probabilities, resulting in such an intermediate effect.

\begin{table}[h]
\centering 
\caption{Correct patches from different search space} 
\renewcommand{\arraystretch}{0.7}
\setlength{\tabcolsep}{14pt}
\begin{tabular}{c|ccc} 
\toprule 
Project & CodeLlama & ARJA-e & CodeLlama+ARJA-e* \\ 
\midrule 
Chart & 8 & 8 & 8 \\
Lang & 18 & 9 & 14 \\
Math & 27 & 17 & 21 \\
Time & 3 & 0 & 2 \\
\midrule 
Total Correct & 56 & 34 & 45 \\
\midrule 
Correct Percentage & 25\% & 15.18\% & 20.09\%\\
\bottomrule 
\end{tabular}
\begin{tablenotes}
        \footnotesize
        \item[*] * refers to the result of searching merged space from both CodeLlama and ARJA-e.  
\end{tablenotes}
\label{differentsource}
\end{table}

\begin{figure}[h]
\centerline{\includegraphics[width=0.7\linewidth]{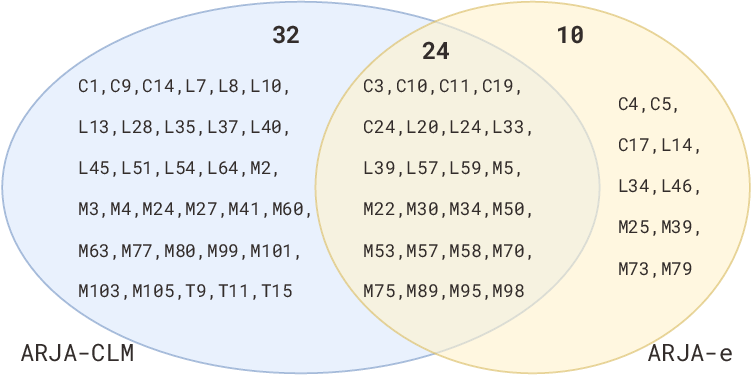}}
\caption{Correct patch Venn diagram of ARJA-CLM and ARJA-e}
\label{vnne}
\end{figure}

 We also notice that among the bugs correctly fixed by ARJA-e, some of them cannot be correctly fixed by ARJA-CLM using CodeLlama, as shown in Figure \ref{vnne}. It indicates that since the search space from ARJA-e and space from CodeLlama do not overlap completely, the bugs that can be fixed by them are differentiated. Therefore, we believe that combining the repair approach of ARJA-CLM and ARJA-e can improve the repair effect to some extent. But in the above analysis, we find that when there are already many candidate sentences for each LBS, combining the search spaces before searching does not improve the repair effect. So to improve the repair effect, we should search different spaces for patches separately and then merge the final patches rather than merge the spaces before searching.

\begin{table}[h]
\centering 
\caption{Average evaluation cost for finding plausible patches} 
\renewcommand{\arraystretch}{0.7}
\setlength{\tabcolsep}{14pt}
\begin{tabular}{c|ccc} 
\toprule 
Project & CodeLlama & ARJA-e & CodeLlama+ARJA-e \\ 
\midrule 
Chart & 196.33 & 288.42 & \textbf{192.50} \\
Lang & \textbf{138.78} & 387.72 & 301.72  \\
Math & \textbf{146.84} & 199.60 & 156.84\\
Time & 274.50 & \textbf{51.00} & 164.00 \\
\midrule 
Overall Average Cost& \textbf{159.19} & 272.49 & 210.35 \\
\bottomrule 
\end{tabular}
\label{evalcost}
\end{table}

\textbf{The impact on search efficiency}. To verify the impact of different search spaces on search efficiency, we collect the intersection of plausible repairs obtained from different search spaces, containing repairs for 57 buggy projects. For each repair, we record the evaluation cost (i.e., the number of fitness evaluations required to find the first plausible patch) and average the costs across repairs for all buggy projects. As shown in Table \ref{evalcost}, except for Time, the evaluation cost for the search space generated by CodeLlama is lower (Lang, Math) or slightly higher (Chart) than other search spaces, and the overall cost is significantly lower than other search spaces. Notably, the Time in the intersection only contains two bugs, which is not representative. Therefore, compared to space generated by ARJA-e, the space by CodeLlama is of higher quality and requires less search cost to obtain patches. Moreover, the average evaluation cost of Codellama+ARJA-e (210.35) is approximately half of that of Codellama (159.19) and ARJA-e (272.49), aligning with the previously observed intermediate effect regarding the number of correct patches.

\subsection{RQ3.The effectiveness of the context-aware prompt}
To explore the effectiveness of the context-aware prompt, for each LBS, we generate prompts with and without extra information. We provide these two different prompts to CodeLlama and obtain two sets of candidate statements. We perform multiobjective evolution on these two different search spaces and obtain the number of plausible and correct patches in Table \ref{woinfo}.

\begin{table}[!htbp]
\centering 
\caption{Correct/plausible patches w/o extra info in prompt} 
\renewcommand{\arraystretch}{0.7}
\begin{tabular}{c|cc} 
\toprule 
Project & With extra info in prompt& Without extra info in prompt\\ 
\midrule 
Chart & 8/14 & 5/12 \\
Lang & 18/28 & 19/34 \\
Math & 27/46 & 20/45 \\
Time & 3/6 & 2/4 \\
\midrule 
Total Plausible/Correct & 56/94 & 46/95 \\
\bottomrule 
\end{tabular}
\label{woinfo}
\end{table}

We find that after adding extra information to the prompt, ARJA-CLM can fix 10 or 21.74\% more bugs, which preliminarily shows that adding extra information to the prompt can improve the repair effect. To further prove the effectiveness, we list in Table \ref{patchusingextra} the correct patches that use the extra information and those unfixable if not adding extra information to prompt.

\begin{table}[!htbp]
\centering 
\caption{Correct patches using extra info} 
\begin{tabular}{cc} 
\toprule 
Correct patches using extra info & Unfixed bugs if without extra info in prompt\\ 
\midrule 
\makecell[l]{C3,L13,L57,M5,M24,M27,\\
M53,M58,M63,M70,M75,T9} & \makecell[l]{C3,L57,M24,M27,M53,M58,T9} \\
\midrule
\multicolumn{1}{l}{$\sum=12$} & \multicolumn{1}{l}{$\sum=7$} \\
\bottomrule 
\end{tabular}
\label{patchusingextra}
\end{table}

\begin{figure}[htbp]
\centering
\subfigure[Example of L57]{\includegraphics[width=0.4\textwidth]{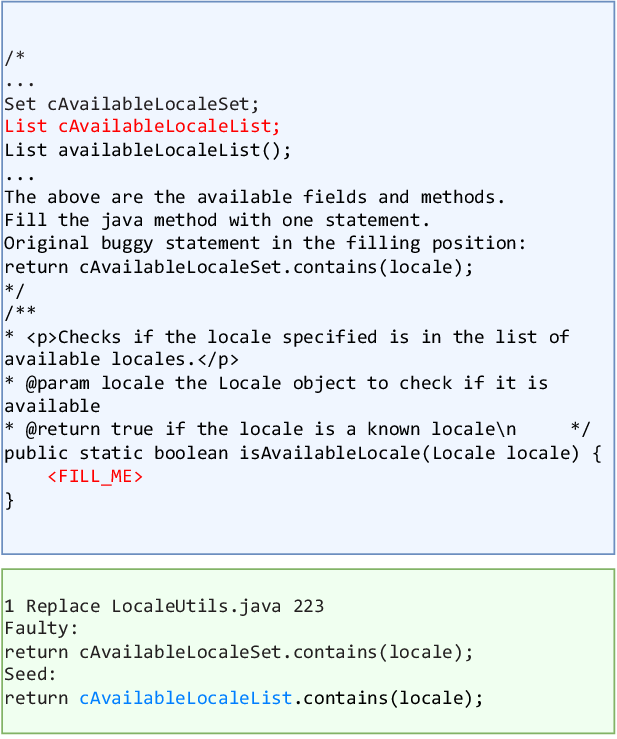}}
\subfigure[Example of C3]{\includegraphics[width=0.4\textwidth]{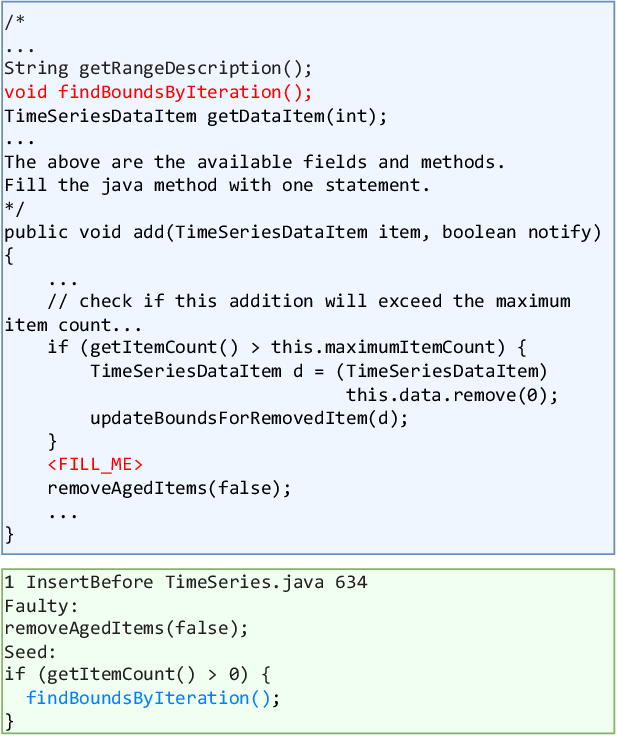}}
\caption{Example of patches using extra info}
\label{patchusinginfo}
\end{figure}

It can be seen that 12 of the correct patches contain fields and methods of the class where the LBS is, and 7 are unfixable bugs without extra info. This shows that CLM can use the field and method information in the comments for code filling, which improves the repair effect to some extent. Figure \ref{patchusinginfo} shows an example of the correct patch generated by CLM using the field and method information for L57 and C3 respectively, where the blue part is the prompt and the green part is the corresponding patch.

\subsection{RQ4.Repair performance of multilocation bugs}

Since ARJA-CLM adopts a multiobjective evolutionary algorithm to enable the CLM-based APR to fix multilocation bugs, we hope to explore the effectiveness of ARJA-CLM in multilocation bug repair.

To compare the performance of ARJA-CLM and ARJA-e in repairing multilocation bugs, considering that ARJA-CLM using CodeLlama has a better repair effect, we selected the multilocation defects repaired by it and those by ARJA-e. Note that as long as the human-written patch of the defect is multi-edited, we consider the defect to be a multilocation defect. 

\begin{table}[htbp]
\centering 
\normalsize
\caption{Fixed multilocation bugs} 
\setlength{\tabcolsep}{12pt} 
\renewcommand{\arraystretch}{0.6}
\begin{tabular}{c|cccc} 
\toprule 
\multirow{2}*{\small{Project}} & \multicolumn{2}{c}{\small{ARJA-CLM}} & \multicolumn{2}{c}{\small{ARJA-e}} \\
\cmidrule(lr){2-3} \cmidrule(lr){4-5}
~ & Plausible & Correct & Plausible & Correct \\
\midrule
Chart & 6 & 2 & 6 & 2 \\
Lang & 10 & 7 & 10 & 4 \\
Math & 20 & 7 & 16 & 4 \\
Time & 2 & 1 & 0 & 0 \\
\midrule
Total & 38 & 17 & 32 & 10 \\
\bottomrule 
\end{tabular}
\label{multilocation}
\end{table}

As shown in Table \ref{multilocation}, among the 120 multi-point defects in the Defect4j benchmark, ARJA-CLM can generate plausible patches for 38 multilocation bugs, of which 17 patches are correct, while ARJA-e can generate plausible patches for 32 multilocation bugs, of which only 10 are correct. Therefore, with the candidate statements from CodeLlama, the search space is more likely to contain patches, making ARJA-CLM fix more multilocation bugs.

\subsection{RQ5.Performance of ARJA-CLM}
\textbf{Results on Defects4J}. To evaluate the performance of our approach, we obtained the plausible patches and correct patches by ARJA-CLM using CodeLlama. Since learning-based APR DEAR \cite{li2022dear} aims at fixing multilocation bugs and also adopts not perfect fault localization, we choose DEAR as the comparison technique.

\begin{table}[!t]
\centering 
\caption{Baseline comparisons} 
\label{comparison}
\renewcommand{\arraystretch}{0.7}
\resizebox{\textwidth}{!}{
\begin{tabular}{c|ccccccccc} 
\toprule 
Project & ARJA-CLM & DEAR & HERCULES & Recoder & DLFix & AlphaRepair* & GAMMA* & TBar & SimFix\\ 
\midrule 
Chart  & 8 & 8 & 8 & 8 & 5 & 9 & \textbf{11} & 9 & 4\\
Lang & \textbf{18} & 8 & 10 & 9 & 5 & 13 & 16 & 5 & 14\\
Math & \textbf{27} & 20 & 20 & 15 & 12 & 21 & 25 & 18 & 9\\
Time & \textbf{3} & \textbf{3} & \textbf{3} & 2 & 1 & \textbf{3} & \textbf{3} & 1 & 1 \\
\midrule 
Total Correct & \textbf{56} & 39 & 41 & 34 & 23 & 46 & 55 & 33 &28\\
\bottomrule 
\end{tabular}}

\begin{tablenotes}
        \footnotesize
        \item[*] * indicates that this column is the result under perfect fault localization  
\end{tablenotes}

\end{table}

As shown in Table \ref{comparison}, ARJA-CLM correctly fixes 56 defects in four defect projects, Chart(8), Lang(18), Math(27), and Time(3), surpassing all repair approaches. In the Lang and Math projects, compared to other approaches, ARJA-CLM fixes the most defects. Compared to DEAR and HERCULES, which can also repair multilocation defects, ARJA-CLM repairs 17 and 15 more defects respectively(i.e., 43\% and 36\% relative improvements). And compared to GAMMA which adopts perfect fault localization, ARJA-CLM using not perfect fault localization still has subtle advantages.
\begin{figure}[h]
\centerline{\includegraphics[width=0.5\linewidth]{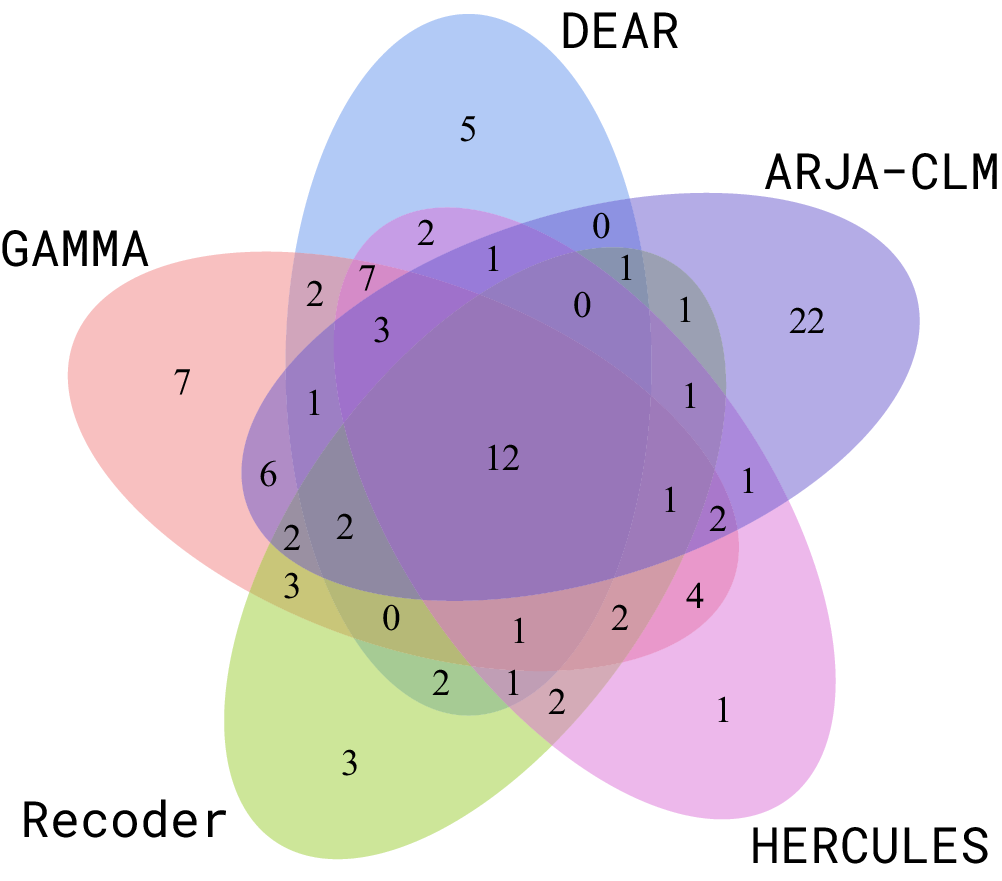}}
\caption{ Venn diagram of correctly repaired bugs}
\label{comparevenn}
\end{figure}
In addition, ARJA-CLM also has some advantages over other state-of-the-art repair approaches, reflecting the feasibility of applying CLM to APR and the potential of introducing traditional repair approaches to CLM-based APR. To further show the performance of ARJA-CLM, we count the number of unique bugs that only ARJA-CLM can fix. As shown in Figure \ref{comparevenn}, ARJA-CLM can fix the most number of unique bugs of 22.

Interestingly, by analyzing the correct patches by ARJA-CLM, we find that the CLM could use the Javadoc comments in prompt to understand the execution process of the method and perform better mask prediction. As shown in Figure \ref{prompt}, since our model prompt included the entire defective method, the Javadoc comments were also added to the prompt. As shown in Figure \ref{javadoc}, for L35 that other APR \cite{b5,b10,b9,b7,b6,b8} could not fix, ARJA-CLM could give the correct patch. The patch throws an exception when both $array$ and $element$ are null, which was clearly stated in the Javadoc comments. We note here that L35 is a multilocation bug and we only present one modification as another is similar.  

\begin{figure}[htbp]
\centering
\begin{lstlisting}[linebackgroundcolor={%
 \ifnum\value{lstnumber}=20
                \color{red!35}
        \fi
           \ifnum\value{lstnumber}=21
                \color{green!35}
        \fi
            \ifnum\value{lstnumber}=9
                \color{yellow!35}
        \fi
}]
// ArrayUtils.java
/**
...
* @param array  the array to "add" the element to, may be <code>null</code>
* @param element  the object to add, may be <code>null</code>
* @return A new array containing the existing elements plus the new element
* The returned array type will be that of the input array (unless null),
* in which case it will have the same type as the element.
* If both are null, an IllegalArgumentException is thrown
* @since 2.1
* @throws IllegalArgumentException if both arguments are null
*/
public static <T> T[] add(T[] array, T element) {
    Class<?> type;
    if (array != null){
        type = array.getClass();
    } else if (element != null) {
        type = element.getClass();
    } else {
-       type = Object.class;
+       throw new IllegalArgumentException("Arguments cannot both be null");
    }
    ...
}
\end{lstlisting} 
\caption{Correct patch generated by ARJA-CLM for bug L35.}
\label{javadoc}
\end{figure}

\textbf{Results on APRCOMP-2024 AI Generated Code Track (Java)}.
Considering the possibility of the Defect4J leaking into the training data of CodeLlama, we further conduct experiments on the newly released benchmark from  APRCOMP-2024 to verify the generalizability of ARJA-CLM. As shown in Table \ref{aprcomp}, out of 100 bugs, ARJA-CLM plausibly repairs 19 bugs and correctly repairs 7 bugs, surpassing all other participating repair tools. This suggests that the repair effectiveness of ARJA-CLM is relatively unaffected by data leakage. Notably, compared to ARJA-e using traditional methods to prepare candidates, ARJA-CLM shows significant improvement in fixing incorrect solutions to the neighboring-bitwise-xor problem available on Leetcode\footnote{\url{https://leetcode.com/problems/neighboring-bitwise-xor/description/}}. As shown in Figure \ref{neighbor-4}, since the erroneous code in the benchmark pertains to algorithmic issues, to replace the return statement "return true", traditional methods struggle to generate candidates by copying redundant code from other parts of the program or using manually defined templates with limited generalizability. In contrast, the CLM can generate correct candidates based on the context of the given problem and erroneous solution. However, we also found limitations in directly using the CLM to generate candidates, especially when fixing incorrect algorithmic solutions which requires multiple modifications to be fixed. Due to the tight data computation flow in algorithmic solutions, the direct context of the single buggy statement is more likely to contain errors besides the erroneous statement itself, misleading the CLM to generate incorrect candidates. In the future, we expect to explore methods that link multiple erroneous statements of a single bug and enable the model to modify multiple erroneous statements simultaneously.
\begin{table}[!t]
\centering 
\caption{Comparison on APRCOMP-2024 benchmark} 
\label{aprcomp}
\renewcommand{\arraystretch}{0.7}
\resizebox{\textwidth}{!}{
\begin{tabular}{c|cccccccc} 
\toprule 
Repairs & ARJA-CLM & ARJA-e & ET & RepairLLAMA &APRER & LLMR & TBar &  ARJA\\ 
\midrule 
Plausible & \textbf{19} & 5 & 4 &1& 1 & 0 & 0  & 0 \\
Correct & \textbf{7} & 5 & 4 &1 & 1 & 0 & 0  & 0 \\
\bottomrule 
\end{tabular}}
\begin{tablenotes}
        \footnotesize
        \item[*] Plausible: passing the public test suite of the defect project.
        \item[*] Correct: passing the private test suite invisible to the repair tool.
\end{tablenotes}
\end{table}

\begin{figure}[htbp]
\centering
\begin{lstlisting}[linebackgroundcolor={%
 \ifnum\value{lstnumber}=5
                \color{red!35}
        \fi
           \ifnum\value{lstnumber}=6
                \color{green!35}
        \fi
        \ifnum\value{lstnumber}=9
                \color{green!35}
        \fi
        \ifnum\value{lstnumber}=10
                \color{green!35}
        \fi
        \ifnum\value{lstnumber}=11
                \color{green!35}
        \fi
}]
class Solution {
    public boolean doesValidArrayExist(int[] derived) {
        int n = derived.length;
        if (n == 1) {
-           return true;
+           return derived[0] == 0;
        }
        int originalFirst = 0;
+       if (derived[0] != 0) {
+           originalFirst=1;
+       }
        for (int i = 0; i < n - 1; i++) {
            originalFirst ^= derived[i];
        }
        return originalFirst == (derived[n - 1] ^ derived[0]);
    }
}
\end{lstlisting} 
\caption{Correct patch generated by ARJA-CLM for neighboring-bitwise-xor problem}
\label{neighbor-4}
\end{figure}

\section{Related work}
\textbf{Search-based APR} \cite{b52,b44,b22,b14,b53,b35,b36,b8}. The approach first prepares a search space of candidate statements based on manually defined templates, statement redundancy assumptions, etc. Then it uses the heuristic search to find patches in the search space. To improve its performance, researchers usually focus on optimizing the search space and the search algorithm. For instance, to optimize the search space, SimFix \cite{b8} believes that the code in the same project has better reference value so it searches for similar code in the project by using code structure and code semantic features. To improve the search algorithm, CapGen \cite{b55} analyzes the context information of mutation operators and repair components, and prioritizes those candidate patches that are more likely to be correct. However, the search space obtained by the traditional search-based APR has a low probability of containing patches. To solve the problem, ARJA-CLM leverages CLMs to generate high-quality candidate statements, which helps the backend search algorithm to find patches.

\textbf{CLM-based APR} \cite{b56,b2,b57,b60,b61,b62,b65}. AlphaRepair \cite{b2} is a typical CLM-based APR that directly generates patches using CLM. For each defect, AlphaRepair extracts the context of the defect location and adds the defect line itself as a comment to the prompt. Then AlphaRepair uses repair templates to replace various syntactic components in the defect line with special masks, resulting in different prompts. Finally, AlphaRepair inputs these prompts into CodeBERT \cite{b13}, and obtains the code components generated by CodeBERT to replace the original buggy components to get patches. Moreover, Appt \cite{b57} improves the accuracy of patch correctness prediction by pre-training and fine-tuning pre-trained models. GAMMA \cite{b56} uses mask patterns generated with manually defined templates in traditional repair approaches such as TBar \cite{b7} to build prompts. However, current CLM-based APR \cite{b2,b56} only focuses on single-point repair while many real-world bugs require multi-point modifications to be fixed, which greatly limits the scope of repairable bugs. To solve the problem, ARJA-CLM introduces multiobjective evolutionary algorithm into CLM-based APR, enabling CLMs to repair multi-point defects.

\textbf{Learning-based APR} \cite{b20,b58,b59,b63,b19,b45,b64}. Learning-based APR aims to tackle program repair problems by leveraging neural networks. Traditional learning-based APR employs neural models to learn repair templates. Recently, powerful NMT-based APR approaches \cite{b19,b45}, often constructed with sequence-to-sequence learning \cite{b24}, treat software repair as a Neural Machine Translation \cite{b27} task. For instance, SequenceR \cite{b20} splits buggy code lines into token sequences. Recoder \cite{b5} utilizes abstract syntax trees of code fragments as input to comprehend code structure. CURE \cite{b19} designs a new code-aware search strategy, which focuses on finding more correct repairs by paying attention to compilable patches and patches with lengths close to the faulty code. Moreover, deep learning can also be used for detecting multiple repair hunks. For example, DEAR \cite{li2022dear} fine-tunes a pre-trained BERT model to learn the repair relationships between multiple statements, and then leverages the sentence pair classification ability of the BERT model to determine whether two statements need to be repaired together.

\section{Threats to validity}
\textbf{Internal}. One internal threat comes from manually verifying the correctness of plausible patches though it is widely used in the field of automated program repair. Due to the limitations of knowledge and the complexity of defects, we may misjudge the correctness of the patch, resulting in experimental bias. To minimize the experimental bias, we spent a lot of time understanding the causes of defects according to the human-written patch as accurately as possible to verify the plausible patches by ARJA-CLM.

Another internal threat to validity is data leakage as part of the training data for CodeLlama and InCoder comes from Github, which means that the correct code for the defect projects may also be used as training data for CLM. However, with prompts containing extra fields and methods that make the prompt different from training data, the repair effect of ARJA-CLM improves, indicating that the experimental results are less affected by model overfitting. Additionally, the buggy programs in the newly released APR-2024 AI Track benchmark are written by GPT, with less data leakage issues. ARJA-CLM also performed well in this test, indicating its performance less affected by data leakage problems.

In addition, another internal validity threat is that different experimental settings may lead to different results. For example, if the search time threshold of the evolutionary algorithm is longer, it is more likely to obtain plausible patches; and for the same prompt input, the output code sequence of the CLMs may vary on different GPUs. Since our experimental results for comparison come directly from the papers whose experimental settings are different, this may affect the fairness of the repair effect comparison. In addition, as shown in Figure \ref{prompt}, the prompt contains the whole method and additional field and method information. To meet the memory capacity limit, we have to trim the prompt, thus losing some important information. Moreover, to minimize the impact of trimming, the token number threshold of the prompt needs to be adjusted according to the memory capacity, making the quality of the candidate sentences from CLM different under different experimental conditions. To solve the above problems, we need to conduct experiments on all repair approaches under the same experimental conditions, which will consume a lot of time.

\textbf{External}. One external threat is that our repair approach is aimed at passing the JUnit test suite in Java programs, for which it cannot generalize well to other programming languages or the Java program without test suites.

Another external threat is that the 224 bugs in our experimental subject Defects4J benchmark cannot fully cover all the defects in the real world. Further experiments are needed on other Java defect datasets \cite{b66,b67,b68} to verify the repair effect of ARJA-CLM.
\section{Conclusion}
In this study, we introduce ARJA-CLM, an APR that applies multiobjective evolution to the CLM-based repair approach, which enables CLMs to repair multilocation bugs. Additionally, we present a novel context-aware prompt generation strategy for each LBS, which incorporates callable fields and methods at the buggy location into the prompt, enabling CLMs to generate more precise candidate statements with the extra information. We evaluate ARJA-CLM on the Defect4J benchmark. The results show that ARJA-CLM can correctly repair more multilocation defects than ARJA-e and the context-aware prompt helps CLMs to generate higher-quality statements. Compared to DEAR and other state-of-the-art repairing approaches, ARJA-CLM repairs more defects. Additionally, to mitigate the impact of dataset leakage, we conduct experiments on the newly released APR-2024 competition benchmark and find that ARJA-CLM also performed excellently, demonstrating its generalizability. Furthermore, we find that the repair effect of ARJA-CLM improves as the quantity of model parameters increased. Notably, ARJA-CLM achieves more bug fixes when employing CodeLlama over InCoder, suggesting that the overall model performance significantly influences the repair performance.

\bibliographystyle{elsarticle-num-names}
\bibliography{refer}

\end{document}